\pdfoutput=1
\documentclass[runningheads]{llncs}
\usepackage[T1]{fontenc}
\usepackage{graphicx}
\usepackage{multirow}
\usepackage{adjustbox}
\begin{document}
\title{Universal Adversarial Framework to Improve Adversarial Robustness for Diabetic Retinopathy Detection}
\titlerunning{Univ. Adv. Framework to improve Adv. Robustness to DR detection}
%
\author{Samrat Mukherjee\inst{1} \and
Dibyanayan Bandyopadhyay\inst{1} \and
Baban Gain \inst{1}\and
Asif Ekbal\inst{1}}
\authorrunning{S. Mukherjee et al.}
\institute{Indian Institute of Technology Patna, Patna, India}
%
\maketitle              
\begin{abstract}
Diabetic Retinopathy (DR) is a prevalent illness associated with Diabetes which, if left untreated, can result in irreversible blindness. Deep Learning based systems are gradually being introduced as automated support for clinical diagnosis. Since healthcare has always been an extremely important domain demanding error-free performance, any adversaries could pose a big threat to the applicability of such systems. 
In this work, we use Universal Adversarial Perturbations (UAPs) to quantify the vulnerability of Medical Deep Neural Networks (DNNs) for detecting DR. To the best of our knowledge, this is the very first attempt that works on attacking complete fine-grained classification of DR images using various UAPs. Also, as a part of this work, we use UAPs to fine-tune the trained models to defend against adversarial samples. We experiment on several models and observe that the performance of such models towards unseen adversarial attacks gets boosted on average by $3.41$ Cohen-kappa value and maximum by $31.92$ Cohen-kappa value. The performance degradation on normal data upon ensembling the fine-tuned models was found to be statistically insignificant using t-test, highlighting the benefits of UAP-based adversarial fine-tuning.

\keywords{Diabetic Retinopathy \and Universal Adversarial Perturbation \and Adversarial training}
\end{abstract}
\section{Introduction}
The global prevalence of diabetes among individuals aged $20$-$79$ was reported to be $460$ million according to the International Diabetes Federation (IDF) in $2019$. Diabetes is associated with serious health issues such as Diabetic Retinopathy (DR), heart attack, and renal failure. Deep learning (DL) systems for DR detection aim to expedite treatment by reducing human effort. Manual detection of DR is time-consuming, requiring trained clinicians to examine each fundus (i.e. the inside, back surface of the eye) image and identify vascular abnormalities associated with the disease.\cite{atlas2015international}

Adversarial attacks pose a significant concern when applying DL-based systems in safety-critical applications. Adversarial examples visually appear similar to normal examples but can mislead trained models into making incorrect classifications. Deploying vulnerable models in healthcare applications can have detrimental consequences. In this paper, we specifically focus on DR and develop adversarially robust methodologies for its detection. Although various AI-based systems have been developed for DR, their poor performance on out-of-distribution samples limits their applicability in real-world scenarios. Therefore, it is crucial to ensure adversarial robustness when developing safety-critical systems.

Adversarial training is a popular approach for improving model robustness against adversarial examples. The proposed method incorporates adversarial examples into the training process, enabling the model to encounter adversarial samples and achieve robust performance on adversarial test samples. PGD-based adversarial training, such as the method proposed by Madry et al. \cite{madry2018towards}, is widely used. However, this method relies on input images and requires the use of target model parameters. Consequently, if the model encounters an unseen perturbation, its performance can significantly deteriorate, making it unsuitable for adversarial situations in the healthcare domain, where it is necessary for the models to be robust to adversarial inputs.


To address these challenges, we propose the use of Universal Adversarial Perturbations (UAPs) \cite{moosavi2017universal} for adversarial training. UAPs generate perturbations that are agnostic to specific images and have demonstrated generalization capabilities across different models. This universal nature assists the model in learning to defend against attacks from unseen perturbations. Our hypothesis is based on the observation that universal perturbations exploit geometric correlations in different parts of the classifier's decision boundary \cite{moosavi2017universal}.

Our contributions as proposed in this paper are as follows:
\begin{enumerate}
  \item Developing a model agnostic adversarial training methodology for robust grading of DR images.
  \item The proposed methodology is task agnostic, so it can be applied to other domains as well.
  \item To the best of our knowledge, this is the first system that robustly classifies DR images on all five grades, rather than just two classes of \textit{DR} and \textit{No-DR}, indicating whether the patient is suffering from the disease or not respectively.
\end{enumerate}

\section{Related Work}

The evaluation of existing deep learning models looks at generalizability and overfitting but does not go far enough into model sensibility and fragility to changes in input. For the first time, the robustness of deep
learning networks in medical imaging were questioned and investigated for state-of-the-art network by utilizing adversarial examples in \cite{10.1007/978-3-030-00928-1_56}. 

To this purpose, the authors analyse various architectures including Inception V3 \cite{https://doi.org/10.48550/arxiv.1512.00567}, Inception V4 \cite{https://doi.org/10.48550/arxiv.1602.07261}, and MobileNet \cite{https://doi.org/10.48550/arxiv.1704.04861}  for classifying skin lesions using FGSM\cite{https://doi.org/10.48550/arxiv.1412.6572}, Deep-Fool\cite{https://doi.org/10.48550/arxiv.1511.04599}, Saliency map attacks, and gaussian noise addition.
It was found that the addition of Gaussian noise merely decreased classification confidence, although practically all adversarial samples had high confidence in being erroneously categorised. As a result, adversarial examples, rather than noisy test images are better suitable for evaluating model robustness. It was concluded that Inception V4 is better for medical classification tasks due to equivalent generalizability and resilience among others, despite any variances in accuracy when different assaults are applied. 
Two models, Inception-ResNet-v2\cite{https://doi.org/10.48550/arxiv.1602.07261} and NasNet-large \cite{https://doi.org/10.48550/arxiv.1707.07012} were extensively analysed in \cite{10.1007/978-3-030-02628-8_10} for chest X-ray classification in various disease categories. These two neural networks were attacked with ten different adversarial attacks. They found that attacks based on white-box gradients were the most effective at deceiving. This could be the rationale behind the preference for gradient-based assaults like FGSM\cite{https://doi.org/10.48550/arxiv.1412.6572} when determining the network's susceptibility. Additionally, research has demonstrated that average-pooling captures more global information than max-pooling, making the neural network more resistant to attacks.
All of the earlier investigations used input-dependent adversarial attacks, which means that each image misclassification employs a unique adversarial perturbation. 
A single perturbation by UAP was created in \cite{hirano2021universal} to cause classification networks to perform poorly. The authors focused on tasks for skin cancer, DR, and pneumonia in two class classification problem of "affected" or "not-affected" by the disease. They demonstrated how adversaries may trick DNNs with minor UAP attacks more quickly and cheaply in terms of memory and time.

\section{Methodology}
This paper introduces a simple approach to training deep learning models specifically for the robust detection of DR against unseen adversarial perturbations. The proposed system is an ensemble of various state-of-the-art image classifiers which were pre-trained to classify given \textit{fundus} images into their correct DR stage.\par
To develop this approach, we first train each of the models individually to classify fundus images correctly. For developing the system, we take seven SOTA image classifiers, namely ConvNextTiny\cite{https://doi.org/10.48550/arxiv.2201.03545}, DenseNet121\cite{https://doi.org/10.48550/arxiv.1608.06993}, EfficientNetB0, EfficientNetB4\cite{https://doi.org/10.48550/arxiv.1905.11946}, MobileNetV4\cite{https://doi.org/10.48550/arxiv.1801.04381}, RegNet\cite{https://doi.org/10.48550/arxiv.2101.00590}, ResNet18\cite{https://doi.org/10.48550/arxiv.1512.03385}, all of which have been applied in several applications like \cite{Zhou2022,Gao2021}. The proposed training process is divided into two stages, \textit{viz.} i) Pre-training followed by Fine-tuning and ii) Adversarial training. We describe these processes elaborately in Section \ref{ptft} and Section \ref{at} respectively.


\subsection{Pre-training \& Fine-tuning}\label{ptft}
The scarcity of data poses a significant challenge in utilizing deep learning models in healthcare. To address this, transfer learning is commonly employed, where a pre-trained model on a different dataset (source domain) is fine-tuned on a task-specific dataset (target domain).
In our approach, we initialize the pre-trained models weights using the weights obtained on the ImageNet dataset \cite{5206848}. This technique is widely adopted to improve the performance of downstream image classification tasks across domains, including healthcare \cite{8301998}, \cite{SWATI201934}, crop pest classification \cite{THENMOZHI2019104906}, and pavement distress detection \cite{GOPALAKRISHNAN2017322}.
For the task of Diabetic Retinopathy (DR) image classification, the final classification layer of each model is redefined. Since the dataset consists of five classes representing different stages of DR, we replace the original layer, which has $1000$ nodes for ImageNet classification, with a fully connected layer having five nodes and softmax activation. This modified layer is then trained on the EyePACS2015 dataset \footnote{\url{https://www.kaggle.com/c/diabetic-retinopathy-detection}}  to learn the specifics of DR image classification.



\begin{table}[h!]
\centering
\caption{Percentage of classes in both the datasets}
\label{tab:tab2}
\begin{tabular}{c|c|c|c}
\hline
\textbf{Label} & \textbf{Stage of DR} & \textbf{EyePACS2015}  & \textbf{APTOS2019}    \\ \hline
0 &  No DR  & 73.5 \% & 49.3 \% \\ \hline
1 &  Mild  & 7\%     & 10.1\%  \\ \hline
2 &  Moderate  & 15\%    & 27.3\%  \\ \hline
3 &  Severe  & 2.5\%   & 8\%       \\ \hline
4 &  Proliferative DR  & 2\%     & 5.3\%   \\ \hline
\end{tabular}
\end{table}
Our key motivation in this paper is to develop systems that are useful for the population demography of countries in South East Asia (SEA) like India, Bangladesh, etc. where diabetes is a very common disease. As per the reports of the International Diabetes Foundation(IDF) \cite{IDFSEA}, as of $2021$, $90$ million adults ($20$-$79$) are living with diabetes in the IDF SEA Region, which is estimated to reach $113$ million by $2030$ and $152$ million by $2045$.  $51$\% of the total number of adults living with diabetes in the region are reportedly undiagnosed, highlighting the gravity of the issue.
Keeping that in mind, we fine-tune each of the models on APTOS2019 dataset\footnote{\url{https://www.kaggle.com/c/aptos2019-blindness-detection}}, which Aravind Eye Hospital, India brings together.
To fine-tune the said models, we first freeze all the model layers except the last classification layer. After that, we train the models individually in an end-to-end fashion on the APTOS2019 dataset.
The models are fine-tuned using $4$-fold stratified cross-validation to ensure having enough examples for each class, thus minimizing the dataset class imbalance problem.

\subsection{Adversarial training}\label{at}
Adversarial training is an effective defense strategy against adversarial attacks. It involves incorporating adversarial examples into the training dataset and training the models to enhance their robustness. While PGD adversarial training achieves high accuracy against $L_\infty$ attacks on standard datasets such as MNIST and CIFAR-10, this paper adopts adversarial training with Universal Adversarial Perturbation (UAP) \cite{moosavi2017universal} due to its target model agnostic nature. Additionally, the doubly universal nature of UAP-generated perturbations can help train models to be robust against unseen adversarial attacks. Thus, we individually perform adversarial fine-tuning using universal adversarial perturbations \cite{REN2020346}.
\\\textbf{Adversarial fine-tuning of individual models:}
For adversarial fine-tuning, we partition the training dataset of APTOS2019 into two parts, one to generate the perturbation vector($D_p$) and the other to test the robustness of the models upon attack($D_r$). For each model, we generate a perturbation vector using $D_p$. Since perturbation vector generation using UAP is an iterative algorithm, we continue to iterate over $D_p$ unless and until the fooling ratio is \textgreater{}90\%. The fooling ratio refers to the ratio of the number of data points whose classification labels changed upon adding the perturbation vector. The motivation for having such a high fooling ratio is to ensure that the model is fine-tuned on difficult examples. 

The number of epochs required to achieve the desired condition is presented in Table \ref{tab:tab1}. Following fine-tuning, the final decision of our proposed system is obtained through majority voting ensembles. This approach draws inspiration from bagging \cite{Breiman1996}, where multiple models are trained on different datasets created through bootstrap aggregating. However, in our case, we do not perform bootstrap sampling on the dataset. Instead, we train each model individually using datasets generated by perturbing the original dataset with their own perturbation vectors. This ensures that each model encounters perturbation vectors generated exclusively by itself, enabling an assessment of their performance on unseen adversarial perturbations in real-world scenarios.
To evaluate the models, we analyze the degradation in performance before and after this fine-tuning process. We conduct a two-fold comparison: first, by assessing the performance of test datasets perturbed using the model's own perturbation vectors, and second, by evaluating the models' robustness against perturbation vectors generated by other models. This comparative analysis allows us to measure the individual models' resilience to unseen perturbations.


\begin{table}[]
\begin{center}
\caption{Number of epochs to attain \textgreater{}90\% fooling ratio}
\label{tab:tab1}
\begin{tabular}{c|c|c}
\hline
\textbf{Model Name}         & \textbf{Fooling rate} & \textbf{Achieved at epoch} \\ \hline
ConvNextTiny       & 90.859       & 2                 \\ \hline
DenseNet121        & 93.997       & 1                 \\ \hline
EfficientNet-B0    & 94.406       & 3                 \\ \hline
EfficientNet-B4    & 92.222       & 1                 \\ \hline
MobileNetV3        & 95.088       & 1                 \\ \hline
Regnet             & 91.132       & 2                 \\ \hline
ResNet18           & 90.040       & 5                 \\ \hline
\end{tabular}
\end{center}
\end{table}

\section{Datasets and Experiments}

\subsection{Dataset description}
The APTOS2019 and EyePACS2015 datasets were collected from their respective Kaggle competitions for this study. These datasets consist of images related to diabetic retinopathy (DR), with severity levels rated by clinicians on a scale of 0 to 4. The APTOS2019 dataset contains 3,662 training samples, while the EyePACS2015 dataset includes 35,126 samples. Both datasets exhibit a significant imbalance towards the No-DR class. The dataset composition is summarized in Table \ref{tab:tab2}.



\subsection{Experimental Setup}
To perform our experiments, we develop all the models in Pytorch\cite{NEURIPS2019_9015}, a deep learning library. We used a single NVIDIA A100 GPU with $40$GB of RAM. 
We train all the individual models for a maximum of $25$ epochs, in which we run all the models for $15$ epochs during the pre-training phase and another $10$ epochs for fine-tuning, with an early stopping criteria of $5$ epochs in both the cases.\\
The optimizer for training used was Adam optimizer\cite{https://doi.org/10.48550/arxiv.1412.6980}, and the batch size was taken to be $16$. To evaluate the models, we perform stratified K-fold cross-validation (with $K=4$) owing to the small dataset size and class imbalance.

\subsection{Pre-processing of dataset}
To enhance the uniformity of retinal images for our model's categorization task, we performed the following pre-processing steps: i) Removal of black regions: Images with noticeable black areas around the eyes, caused by differing aspect ratios, were trimmed as they contained irrelevant data. We converted the images to greyscale and identified black areas based on pixel intensity, removing vertical and horizontal black rectangles. ii) Resizing: All images were resized to a standard dimension of 224 x 224 pixels after eliminating black regions. iii) Standardization of eye shape: Circular cropping around the center of each image was applied to achieve a consistent eye shape representation. iv) Smoothing: A Gaussian filter was used to reduce lighting and brightness variations in the pre-processed images.
The results of this set of pre-processing steps are depicted in Fig 1.

\begin{figure}[h]
    \centering
    \includegraphics[scale=0.25]{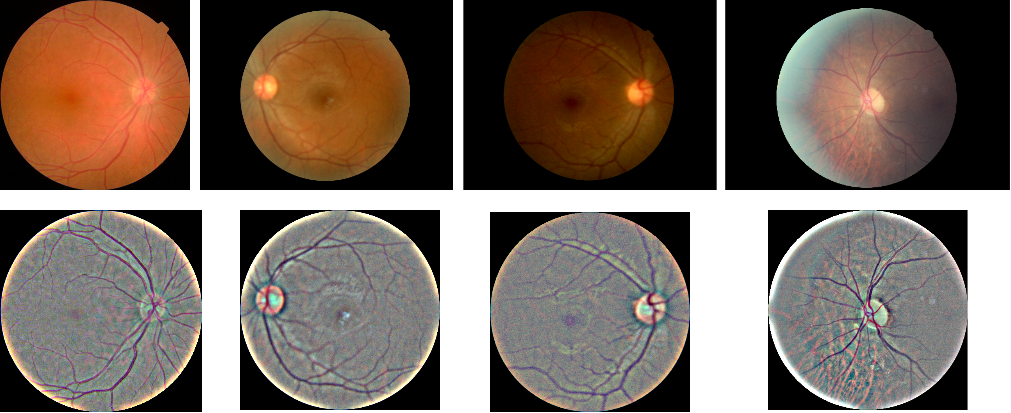}
    \caption{Pre-processing of DR images to uniform shape, size, and illumination. Here, Top(input), Bottom(output)}
    \label{fig:aptosim}
\end{figure}

\subsection{Performance on test set}
We have taken seven SOTA image classifiers for the base models for the ensemble learning strategy. These are ConvNextTiny, DenseNet121, EfficientNetB0, EfficientNetB4, MobileNetV3, RegNet, ResNet18. Individual models are then tested on the test set. The individual performance of the models, as described in Table \ref{tab:tab4}, suggests that the inter-rater similarity of all the individual models and the clinician is "Almost perfect" as depicted in Table \ref{tab:tab3}. Quadratic Cohen Kappa is taken to be the measure of the performance of the models. It is because it is ubiquitous in the medical domain to have two clinicians grading the same instance differently. Hence, Quadratic Cohen Kappa is considered an aligned metric towards human judgment compared to other metrics like accuracy, precision, recall, etc.

\begin{table}[]
\begin{center}
\caption{Quadratic Cohen Kappa interpretation Table \cite{Landis1977-bq}}
\label{tab:tab3}
\begin{tabular}{c|c}
\hline
\textbf{Quadratic Cohen-Kappa}    & \textbf{Level of Agreement} \\ \hline
\textgreater{}0.8 & Almost perfect              \\ \hline
\textgreater{}0.6 & Substantial                 \\ \hline
\textgreater{}0.4 & Moderate                    \\ \hline
\textgreater{}0.2 & Fair                        \\ \hline
\textgreater{}0   & Slight                      \\ \hline
\textless{}0      & No agreement                \\ \hline
\end{tabular}
\end{center}
\end{table}

\begin{table}[]
\begin{center}
\caption{Test set results of individual models}
\label{tab:tab4}
\begin{tabular}{c|c}
\hline
\textbf{Model Name}       & \textbf{Quadratic Cohen Kappa} \\ \hline
ConvNextTiny     & 0.8613                \\ \hline
DenseNet121      & 0.8466                \\ \hline
EfficientNet B0  & 0.8315                \\ \hline
EfficientNet B4  & 0.8187                \\ \hline
MobileNetV3      & 0.8263                \\ \hline
RegNet & 0.8149                \\ \hline
ResNet18         & 0.8451                \\ \hline
Ensemble of all models         & 0.8554                \\ \hline
Ensemble of except ConvNextTiny         & 0.8492                \\ \hline
\end{tabular}
\end{center}
\end{table}
It is observed that upon ensembling all the models, the Cohen Kappa similarity is better than that of all the individual models except ConvNextTiny. A simple ablation by ensembling all the models except ConvNextTiny was performed. The new ensemble's performance was quantitatively better than all the six individual models. This highlights that ensembling does help in better model development in DR stage classification.  

\subsection{Attacking the models}
The performance of the models with normal examples for all the individual models and their ensembles gives "almost perfect" matchings \ref{tab:tab3}. Now we attack the models using all seven perturbed datasets. We use the already generated perturbations from the individual models and create a separate dataset from each perturbation vector. We attack the models individually with all such datasets. To generate the perturbation vector, as mentioned earlier, the stopping criterion is attained when the fooling ratio is more significant than 90\%. 
\begin{table*}[t!]
\caption{Performance of models before adversarial fine-tuning}
\label{tab:tab5}
\adjustbox{width=\textwidth}{
\begin{tabular}{c|ccccccccc}
\hline
                                                                           & \multicolumn{9}{c}{\textbf{Model name}}                                                                                                                                                                                                                                                                  \\ \cline{2-10} 
                                                                                             & \multicolumn{1}{c|}{}               & \multicolumn{1}{c|}{\textbf{ConvNextTiny}} & \multicolumn{1}{c|}{\textbf{DenseNet121}} & \multicolumn{1}{c|}{\textbf{EfficientNetB0}} & \multicolumn{1}{c|}{\textbf{EfficientNetB4}} & \multicolumn{1}{c|}{\textbf{MobileNetV3}} & \multicolumn{1}{c|}{\textbf{RegNet}} & \multicolumn{1}{c|}{\textbf{ResNet18}} & \textbf{Ensemble} \\ \hline
  & \multicolumn{1}{c|}{\textbf{ConvNextTiny}}   & \multicolumn{1}{c|}{0.2092}       & \multicolumn{1}{c|}{0.3571}      & \multicolumn{1}{c|}{0.4748}         & \multicolumn{1}{c|}{0.2982}         & \multicolumn{1}{c|}{0.1755}      & \multicolumn{1}{c|}{0.2737} & \multicolumn{1}{c|}{0.2718} & \textbf{0.5429}   \\
\cline{2-10}    & \multicolumn{1}{c|}{\textbf{DenseNet121}}    & \multicolumn{1}{c|}{\textbf{0.6120}}       & \multicolumn{1}{c|}{0.2744}      & \multicolumn{1}{c|}{0.2531}         & \multicolumn{1}{c|}{0.0762}         & \multicolumn{1}{c|}{0.1487}      & \multicolumn{1}{c|}{0.6017} & \multicolumn{1}{c|}{0.0290} & 0.4984   
\\ \cline{2-10} \multicolumn{1}{c|} {\textbf{Perturbation}}
                                                                                             & \multicolumn{1}{c|}{\textbf{EfficientNetB0}} & \multicolumn{1}{c|}{0.4250}       & \multicolumn{1}{c|}{0.5031}      & \multicolumn{1}{c|}{0.1343}         & \multicolumn{1}{c|}{0.2911}         & \multicolumn{1}{c|}{0.0860}      & \multicolumn{1}{c|}{\textbf{0.6788}} & \multicolumn{1}{c|}{0.0239} & 0.4755  
\\ \cline{2-10} \multicolumn{1}{c|} {\textbf{vector}}
                                                                                             & \multicolumn{1}{c|}{\textbf{EfficientNetB4}} & \multicolumn{1}{c|}{0.6755}       & \multicolumn{1}{c|}{0.1545}      & \multicolumn{1}{c|}{0.4876}         & \multicolumn{1}{c|}{0.4063}         & \multicolumn{1}{c|}{0.6604}      & \multicolumn{1}{c|}{0.6453} & \multicolumn{1}{c|}{0.2232} & \textbf{0.7255}   
\\ \cline{2-10} \multicolumn{1}{c|} {\textbf{created}}
                                                                                             & \multicolumn{1}{c|}{\textbf{MobileNetV3}}    & \multicolumn{1}{c|}{0.4856}       & \multicolumn{1}{c|}{0.5281}      & \multicolumn{1}{c|}{0.4149}         & \multicolumn{1}{c|}{0.1762}         & \multicolumn{1}{c|}{0.0578}      & \multicolumn{1}{c|}{\textbf{0.7268}} & \multicolumn{1}{c|}{0.0342} & 0.5739   
\\ \cline{2-10} 
                                                                                             & \multicolumn{1}{c|}{\textbf{RegNet}}         & \multicolumn{1}{c|}{0.5526}       & \multicolumn{1}{c|}{0.5082}      & \multicolumn{1}{c|}{0.5254}         & \multicolumn{1}{c|}{0.1583}         & \multicolumn{1}{c|}{0.0512}      & \multicolumn{1}{c|}{0.4165} & \multicolumn{1}{c|}{0.1181} & \textbf{0.5540}   \\ \cline{2-10} 
                                                                                             & \multicolumn{1}{c|}{\textbf{ResNet18}}         & \multicolumn{1}{c|}{0.2640}       & \multicolumn{1}{c|}{0.3764}      & \multicolumn{1}{c|}{0.1759}         & \multicolumn{1}{c|}{0.1218}         & \multicolumn{1}{c|}{0.3391}      & \multicolumn{1}{c|}{0.2826} & \multicolumn{1}{c|}{0.4418} & \textbf{0.5693}   \\ \hline
\end{tabular}%
}
\end{table*}
A difference in the performance of the models was observed when we tested the individual models with the test dataset perturbed by perturbation vectors of the respective model. We perform a statistical significance test (t-test) to quantify the difference in the results.
Assuming the null hypothesis, $H_0$ to be the performance of the models upon seeing adversarial samples does not change, the p-value for the difference of Quadratic Cohen Kappa metric from the normal test dataset to the perturbed test set comes out to be $7.5411e-05$ which is much lesser than $0.05$. Hence, the degradation of the performance is \textbf{statistically significant.}

Each model is further tested on the datasets modified by perturbation vectors of other models. This is shown in Table \ref{tab:tab5}, where the column indicates the model which was being tested, and the rows represent the model whose perturbation vector is used to generate the dataset. We perform an ensemble of these models' predictions by majority voting and then test them. As observed, the performance upon ensembling improves for some of the datasets.

The diagonal entries of Table \ref{tab:tab5} indicate that the model is attacked by adversarial perturbation generated from the model itself. For the case of RegNet, ResNet18, MobileNetV3, EfficientNetB4, and DenseNet121, the performance degradation is more when attacked by a dataset generated by perturbation vectors of ResNet18, DenseNet121, RegNet, DenseNet121, EfficientNetB4 respectively. Among these five, perturbation vectors of EfficientNetB4 and DenseNet121 affect each other more than any other architectures. The ensembling of the models is observed to have helped in four cases among the seven models taken. Hence, it indicates that a system based on the ensemble of models is better than a system based on the individual model only. But the performance of such systems is "Moderate", which is not acceptable in the healthcare domain.

\subsection{Test set performance after adversarial fine-tuning}
We perform adversarial training of the models with the dataset perturbed using perturbation vectors of the model itself. This ensures that the model has been trained on only one type of adversarial perturbation. Taking the example of ResNet18, for example, we adversarially fine-tune ResNet18 architecture using an adversarial dataset that is perturbed by the perturbation vector of ResNet18 only. We do not fine-tune the ResNet18 with any other perturbed dataset. Following fine-tuning, each model is further tested on the datasets, individually modified by the other perturbation vectors.

\begin{table*}[ht!]
\caption{Performance of models after adversarial fine-tuning}
\label{tab:tab6}

\adjustbox{width=\textwidth}{
\begin{tabular}{c|ccccccccc}
\hline
    & \multicolumn{9}{c}{\textbf{Model name}}                                                                              \\ \cline{2-10} 
 & \multicolumn{1}{c|}{}    & \multicolumn{1}{c|}{\textbf{ConvNextTiny}} & \multicolumn{1}{c|}{\textbf{DenseNet121}} & \multicolumn{1}{c|}{\textbf{EfficientNetB0}} & \multicolumn{1}{c|}{\textbf{EfficientNetB4}} & \multicolumn{1}{c|}{\textbf{MobileNetV3}} & \multicolumn{1}{c|}{\textbf{RegNet}} & \multicolumn{1}{c|}{\textbf{ResNet18}} & \textbf{Ensemble} \\ 
 \hline
 
 & \multicolumn{1}{c|}{\textbf{ConvNextTiny}}   & \multicolumn{1}{c|}{\textbf{0.8245}}                & \multicolumn{1}{c|}{0.6257}               & \multicolumn{1}{c|}{0.6119}                  & \multicolumn{1}{c|}{0.7958}                  & \multicolumn{1}{c|}{0.6558}               & \multicolumn{1}{c|}{0.5855}          & \multicolumn{1}{c|}{0.8166}          & 0.7888            
\\ \cline{2-10} \multicolumn{1}{c|} {\textbf{Perturbation}} & \multicolumn{1}{c|}{\textbf{DenseNet121}}    & \multicolumn{1}{c|}{0.7942}                & \multicolumn{1}{c|}{0.8261}     & \multicolumn{1}{c|}{0.8352}                  & \multicolumn{1}{c|}{0.7994}                  & \multicolumn{1}{c|}{0.8043}               & \multicolumn{1}{c|}{0.7731}          & \multicolumn{1}{c|}{0.7073}          & \textbf{0.8387}            
\\
\cline{2-10} \multicolumn{1}{c|} {\textbf{vector}}
& \multicolumn{1}{c|}{\textbf{EfficientNetB0}} & \multicolumn{1}{c|}{0.7998}                & \multicolumn{1}{c|}{0.7463}               & \multicolumn{1}{c|}{0.8212}                  & \multicolumn{1}{c|}{0.7985}                  & \multicolumn{1}{c|}{0.7936}               & \multicolumn{1}{c|}{0.7906}          & \multicolumn{1}{c|}{0.7870}          & \textbf{0.8271}         
\\ 
\cline{2-10} 
 \multicolumn{1}{c|} {\textbf{created}}   & \multicolumn{1}{c|}{\textbf{EfficientNetB4}} & \multicolumn{1}{c|}{0.7515}                & \multicolumn{1}{c|}{0.8155}               & \multicolumn{1}{c|}{0.8013}                  & \multicolumn{1}{c|}{0.8516}                  & \multicolumn{1}{c|}{0.8084}               & \multicolumn{1}{c|}{0.8295}          & \multicolumn{1}{c|}{0.8032}          & \textbf{0.8586 }           \\ \cline{2-10} 
                                                                                                    & \multicolumn{1}{c|}{\textbf{MobileNetV3}}    & \multicolumn{1}{c|}{0.8054}                & \multicolumn{1}{c|}{0.6986}               & \multicolumn{1}{c|}{0.8131}                  & \multicolumn{1}{c|}{0.7521}                  & \multicolumn{1}{c|}{0.8019}               & \multicolumn{1}{c|}{0.7575}          & \multicolumn{1}{c|}{0.8070}          & \textbf{0.8317}            \\ \cline{2-10} 
                                                                                                    & \multicolumn{1}{c|}{\textbf{RegNet}}         & \multicolumn{1}{c|}{0.7847}                & \multicolumn{1}{c|}{0.6371}               & \multicolumn{1}{c|}{0.8241}                  & \multicolumn{1}{c|}{0.7590}                  & \multicolumn{1}{c|}{0.7910}               & \multicolumn{1}{c|}{0.7538}          & \multicolumn{1}{c|}{0.8177}          & \textbf{0.8251}            \\ \cline{2-10} 
                                                                                                    & \multicolumn{1}{c|}{\textbf{ResNet18}}         & \multicolumn{1}{c|}{0.8209}                & \multicolumn{1}{c|}{0.5922}               & \multicolumn{1}{c|}{0.1772}                  & \multicolumn{1}{c|}{0.5899}                  & \multicolumn{1}{c|}{0.6751}               & \multicolumn{1}{c|}{0.6684}          & \multicolumn{1}{c|}{\textbf{0.8256}}          & 0.6451            \\ \hline
\end{tabular}}
\end{table*}

Upon this type of adversarial fine-tuning, we observe that the performance got boosted by a huge margin. Apart from the increase in the Cohen-Kappa value of the principal diagonal entries in Table \ref{tab:tab6}, for example, if we compare the performance of DenseNet121 on the dataset, which is perturbed by perturbation vector of EfficientNetB4, we see that Cohen-kappa value got increased to 0.8155 after fine-tuning, which was earlier 0.1545. A similar performance increase can also be observed in case of RegNet when attacked by images from the dataset, which is perturbed by the perturbation vector of MobileNetV3. The Cohen-kappa value got increased from 0.0512 to 0.7910. This indicates that the models which are adversarially fine-tuned using the UAP as the attacking algorithm are much more equipped to 
to perform against the unseen adversarial attacks. It is also evident that with ensembling, the entire system's performance becomes much more reliable compared to the individual models.\\ 
\subsection{Statistical significance computation before and after adversarial training}
To find the statistical significance of the ensembling of models, we calculate the p-value with a similar assumption that the null hypothesis, $H_0$ to be the performance of the adversarially fine-tuned models upon testing on adversarial samples does not change, the p-value for the difference of Cohen Kappa metric from the normal test dataset to the perturbed test set after fine-tuning comes out to be $0.7610$ which is much higher than $0.05$. Hence, the degradation of performance is \textbf{statistically insignificant}. Although statistically insignificance does not guarantee that the effect is biologically insignificant, there is still a certain degree of reliability this methodology of adversarial training brings in.
The difference in the model's performance on the dataset perturbed by the perturbation vector of itself before and after fine-tuning is also calculated similarly. The p-value of that came as $8.96 \times 10^-5$, indicating that the performance boost is \textbf{statistically significant}.

\section{Conclusion}
In this work, we use the UAP algorithm to perform adversarial training of models for robustly classifying Diabetic Retinopathy images. The key motivation was to ensure the model's performance did not drop in for unseen attacks. We saw that even a simple combination of ensembling of adversarially fine-tuned models using majority voting enhances performance upon fine-tuning using perturbations generated by the UAP algorithm. Upon calculating the statistical significance of the change in the model performance, based on results obtained from models before and after fine-tuning, it can be easily said that the difference between the individual models and also the ensembled performance before and after adversarial fine-tuning using UAPs is statistically significant on both seen and unseen perturbations.

%
%
%
%

\end{document}